\documentclass[journal=jacsat,manuscript=article]{achemso}

\usepackage{achemso}
\usepackage[version=3]{mhchem} % Formula subscripts using \ce{}
\usepackage[T1]{fontenc}       % Use modern font encodings
\usepackage{color}
\usepackage{bm}
\usepackage{soul}
\usepackage{multirow}

\SectionNumbersOn

\definecolor{NavyBlue}{rgb}{0.00,0.00,0.90}
\definecolor{Red}{rgb}{1.00,0.00,0.00}
\definecolor{Green}{rgb}{0.58,0.0,0.83}

\author{Emanuele Coccia}
\affiliation{S3 Center, CNR Institute of Nanoscience, Via Campi 213/A, 41125 Modena, Italy}
\affiliation{Dipartimento di Scienze Fisiche e Chimiche, Universit\'a degli Studi dell'Aquila, via Vetoio, 67100, L'Aquila, Italy}
\email{emanuele.coccia@nano.cnr.it}

\author{Daniele Varsano}
\affiliation{S3 Center, CNR Institute of Nanoscience, Via Campi 213/A, 41125 Modena, Italy}
\email{daniele.varsano@nano.cnr.it}

\author{Leonardo Guidoni}
\affiliation{Dipartimento di Scienze Fisiche e Chimiche, Universit\'a degli Studi dell'Aquila, via Vetoio, 67100, L'Aquila, Italy}
\email{leonardo.guidoni@univaq.it}

\title[Geometry and spectra of oxyluciferin by VMC and MBGFT] {Theoretical S$_1 \leftarrow$ S$_0$ absorption energies of the anionic forms of oxyluciferin by Variational Monte Carlo and Many Body Green's Function Theory}

\begin{document}


\begin{abstract}

The structures of three negatively charged forms (anionic keto-1 and enol-1, dianonic enol-2) of oxyluciferin (OxyLuc), which are the most probable emitters responsible for the firefly bioluminescence, have been fully relaxed at the variational Monte Carlo (VMC) level. Absorption energies of the S$_1 \leftarrow$ S$_0$ vertical transition have been computed using different levels of theory, such as TDDFT, CC2 and many body Green's function Theory (MBGFT).
The use of MBGFT, by means of the Bethe-Salpeter (BS) formalism, on VMC structures provides results in excellent agreement with the value (2.26(8) eV) obtained by action spectroscopy experiments for the keto-1 form (2.32 eV). 
To unravel the role of the quality of the optimized ground state geometry, BS excitation energies have also been computed on CASSCF geometries, inducing a 
non negligible blue shift (0.08 and 0.07 eV for keto-1 and enol-1 forms, respectively) with respect to the VMC ones.
Structural effects have been analyzed in terms of over- or under-correlation along the conjugated bonds of OxyLuc by using different methods for the ground-state optimization. The relative stability of the S$_1$ state for the keto-1 and enol-1 forms depends on the method chosen for the excited state calculation, thus representing a fundamental caveat for any theoretical study on these systems.
Finally, Kohn-Sham HOMO and LUMO orbitals of enol-2 are (nearly) bound only when the dianion is embedded into a solvent (water and toluene in the present work); excited state calculations are therefore meaningful only in presence of a dielectric medium which localizes the electronic density.
The combination of VMC for the ground state geometry and BS formalism for the absorption spectra clearly outperforms  standard TDDFT and quantum chemistry approaches.

\end{abstract}

\section{Introduction}

\begin{figure}[tbp]
\caption{Lewis representation of the keto-1, enol-1, and enol-2 forms of oxyluciferin studied in this work.}
\label{fig1}
\includegraphics[width=0.3\textwidth]{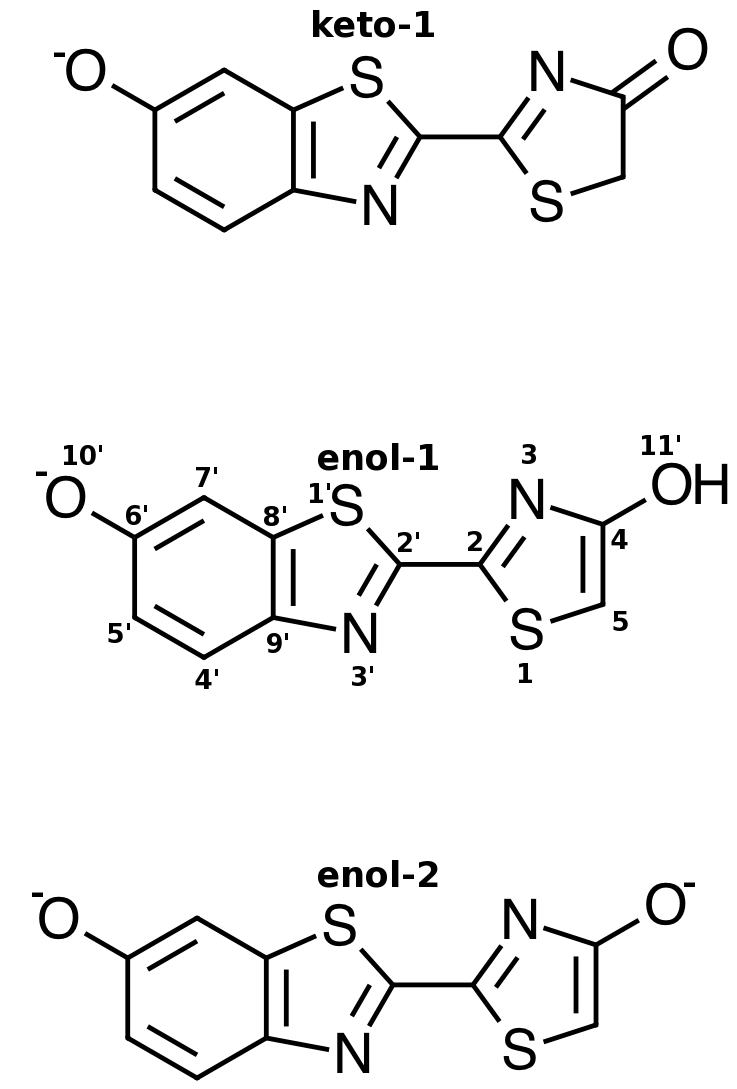}
\end{figure}

Bioluminescence \cite{navi11,hosse11,silva11,silva11a, navi13, uga02} occurs in several living organisms, like fish, insects, algae and bacteria. Animals use bioluminescence for a variety of purposes, such as communication, camouflage, and self-defense. 
Light emission is the product of a reaction catalyzed by the luciferase in presence of ATP, Mg$^{2+}$ and O$_{2}$, in which D-luciferin is transformed into
oxyluciferin (OxyLuc, Figure \ref{fig1}) \cite{fraga08}. The light emitter of the firefly is the OxyLuc in the excited state. \\
A wide debate on the emitting form of OxyLuc produced a large number of computational and experimental studies on the optical properties of keto and enol tautomers and their respective anions \cite{orl03,ren05,yang07,min10,liu08,chen11,sto13}. 
The structure of OxyLuc is the same for various luciferases \cite{woo95,uga08}, but the emitted light naturally ranges from green (530 nm) to red (635 nm) \cite{mc69,viv02}.
It has been argued that several factors affect the OxyLuc bioluminescence color: the pH of the environment \cite{mil10,silva11b}, according to the keto-enol tautomeric and dissociation equilibria of OxyLuc \cite{orl03,ren05,yang07,min10,liu08,chen11}, the distortion of OxyLuc in its excited state\cite{capra94}, the polarization of the micro-environment \cite{deluca69}, the resonance structures of the emitting form \cite{bran02,bran04} and the rigidity of the luciferase binding pocket \cite{naka06}.
Important advances in the comprehension of the mechanisms regulating the bioluminescence mechanism have been achieved in the last ten years also thanks to computational studies, mainly based on multi-scale QM/MM techniques~\cite{naka07,naka09,navi10,min10a,tag09,cai11}. \\
Despite the large amount of studies in the literature, there is still no general consensus on the relative stability of the
tautomers (also with different protonation states) and how the interaction with the external environment modifies the stability of the various forms of OxyLuc.
Several experimental~\cite{bran02,bran04} and theoretical~\cite{navi10, chen11,silva11b,song11,Cheng15} studies indicate the keto-1 form (Figure \ref{fig1}) as the light emitter. 
Nevertheless, Naumov {\it et al.}, reporting the crystal structure of OxyLuc in the enol-1 form (Figure \ref{fig1}), proposed that the enol-1 is the emitter  \cite{nau09,gho15,sil15}. \\
%On one hand, experiments \cite{bran02,bran04}, QM/MM calculations \cite{navi10} and QM investigations \cite{chen11,silva11b,song11,Cheng15} support the hypothesis that the emitter is the keto-1 form (Figure \ref{fig1}) \st{emitting yellow-green light}. \st{Further theoretical studies, using DFT and multireference methods, confirm this hypothesis} \cite{chen11,silva11b,song11,Cheng15}.
%On the other hand, in 2009 Naumov {\it et al.}, reporting the crystal structure of OxyLuc in the enol-1 form (Figure \ref{fig1}), proposed that the enol-1 is the light emitter  \cite{nau09,gho15,sil15}. \\
%Moreover, the yellow-green light found in living organisms could be interpreted as the superposition of different colors coming from the coexistence of the different species in a triple equilibrium \cite{uga05}. \\
% Important contributions to unveil the nature of the bioluminescence mechanism comes from
%the use of analogues of OxyLuc to constrain the ionization of both hydroxyl groups and keto-enol tautomerization permits a systematic understanding of the form of OxyLuc responsible for the emission \cite{reb13,gho15,Cheng15}, since the spectra of the analogues reflect the luminescence properties of the corresponding OxyLuc.\\
In order to understand the photophysics of OxyLuc in complex environments, first a clear understanding of the ground state and absorption properties in the gas phase is needed. 
In this context, gas phase experiments are very useful to capture the intrinsic electronic properties of OxyLuc without any perturbation given by the surrounding environment and to produce a reference analysis for the gas phase absorption \cite{sto13}.
Action spectroscopy on the bare OxyLuc singly-charged anion produces a quite broad absorption band at 548 $\pm$ 10 nm (2.26 $\pm$ 0.08 eV) \cite{sto13}.  
Computational studies \cite{sto13} carried out so far were only partially in agreement with the experimental findings, showing a large blue shift with respect to the experiments: about 70 nm \cite{sto13} for EOM-CCSD  and 58 and 84 nm \cite{sto13} for TDDFT/B3LYP and TDDFT/CAM-B3LYP respectively. \\
%TDDFT  values carried on a structure obtained at B3LYP/Def2-SVPD level of keto-1 \cite{sto13} provide only a partial agreement with the experimental finding, showing a large blue-shift with respect to the experiment: about 70 nm for EOM-CCSD and 58 nm and 84 for TDDFT/B3LYP and TDDFT/CAM-B3LYP respectively.
%\st{Surprisingly in the same study, a good agreement was found using local and semi-local functionals (SVWN and BLYP) probably due to a fortuitous compensation of errors, as also stated by the authors.} \\
Following the indications of Refs \cite{chen11,Cheng15} we study here the absorption properties of the gas phase  keto-1, enol-1 and enol-2 forms of OxyLuc, focusing the attention on the first (bright) excitation, S$_{1} \leftarrow$ S$_{0}$. 
In this work we aim: i) to
study the possible interplay between the optimized ground state geometric parameters and the absorption properties of the OxyLuc forms; ii) to define a computational procedure able to accurately compute the relative shift in the absorption for the three forms; iii) to understand the reliability of gas phase calculations of the dianion enol-2. \\%by looking at the HOMO-LUMO gap. \\
When comparing different methods for the evaluation of the excited state properties of a conjugated molecule it is crucial to start from reference ground state structures, since excitation energies can be highly affected by the fine structure of the bond length pattern. 
%affects the absorption \st{spectra}. 
Recently, we have shown how Quantum Monte Carlo (QMC) \cite{fou+01rmp} methods can provide an accurate  ground state geometry for polyenes and conjugated biochromophores. \cite{bar+15jctc_b,Coccia:2012ex,Coccia14,var14} 
Following these works we have optimized the ground state geometry of the three forms by means of variational Monte Carlo (VMC) using the Jastrow Antisimmetrized Geminal Power ansatz (JAGP) \cite{cas+03jcp,cas+04jcp}. 
The combined use of QMC and JAGP has proven to be successful in several applications of physical and (bio)chemical interest, from small molecules (like methylene) to cobalt-based catalysts for the water splitting reaction \cite{bar+12jctc,coccia12,Coccia:2012kz,bar+12jcp,bar+15jctc_a,bar+15jctc_b,Coccia:2012ex,var14,Coccia14,rev16, zen14, zen15,bar15a,chu16}. 
The JAGP ansatz has been recently extended to the calculation of excitation energies \cite{neu16,zha16a,zha16b} of small molecules and model systems.
The VMC/JAGP ground state structures obtained for the keto-1, enol-1, and enol-2 forms have been then compared with those obtained within the DFT framework using different classes of functionals and with CASSCF(18,15) in Ref~\cite{liu08}.\\
%also optimized at DFT level using different functionals. A detailed comparison with CASSCF(18,15) \cite{liu08} structures has also been done. \\
The excited state calculations  on keto-1 and enol-1 forms have been carried out using the many body Green's function theory (MBGFT) \cite{fetter1971quantum,onida2002electronic,marini2009yambo}, CC2 and TDDFT with several functionals. 
Within MBGFT, the GW approximation and the Bethe-Salpeter (BS) formalism~\cite{hedin1965new,strinati1982dynamical} have been recently  applied to the study of optical properties of gas-phase \cite{ma2009modeling,blase2011charge,baumeier2012frenkel,duchemin2013resonant,korbel2014benchmark,baumeier2014electronic,Coccia14,var14,jacquemin2015assessment} and embedded (in a protein complex environment, for instance see Ref \cite{var14}) molecular systems, showing results in good agreement with the experimental findings. For neutral excitations, the accuracy is comparable to that of high-level wave function methods~\cite{jacquemin20150,bruneval2015systematic,jacquemin2015benchmarking}, but reliable results are also obtained for charge-transfer excitations~\cite{stein2009prediction,duchemin2012short}. 
In a recent work Noguchi {\it et al.} \cite{nog14} applied the GW/BS approach to the ionic form of the luciferin in order to describe Rydberg and resonance excitations. \\
In this work we have also carried out calculations on the dianionic enol-2 form, keeping in mind that, due to the unbound character of the density, ground- and excited-state calculations would not be significant unless solvent effects are considered to stabilize the frontier orbitals \cite{Cheng15,hiy15,nog16}. 
For this reason, we have also performed TDDFT calculations for enol-2 and the other forms coupled to a continuum polarizable model (PCM), describing water and toluene solvents.  \\
%The gas-phase dianionic enol-2 form is characterized by HOMO and LUMO orbitals with continuum-like character, making the corresponding electronic gap extremely sensitive to the chosen basis set. Therefore, ground- and excited-state calculations on the bare ion are therefore seen to be meaningless, unless solvent effects are considered, as the addition of an external medium stabilizes the frontier orbitals \cite{Cheng15,hiy15,nog16}. For this reason, we have also performed TDDFT calculations for enol-2 coupled to a continuum polarizable model (PCM), describing water and toluene solvents.  \\
The present work is organized as follows:  in Section \ref{theo} we briefly describe the theoretical methodology and discuss the computational details. 
In section~\ref{res} we first report the geometrical properties of the studied forms with different approaches and next we discuss the first optical vertical excitations of the keto-1 and enol-1 forms:  our results are systematically compared with the experimental absorption of the bare ion \cite{sto13} and with previous theoretical studies at TDDFT \cite{sto13} and MS-CASPT2 \cite{chen11} levels of theory. Our analysis of the electronic properties of enol-2 is illustrated in  a dedicated subsection.
Finally in Sec.\ref{conclusions} we summarize our findings, underlining the good performance of VMC/BS for the study of the S$_1 \leftarrow$ S$_0$ excitation of the anionic forms of OxyLuc.

\section{Theoretical methods and computational details}
\label{theo}

\subsection{Geometry optimization}
Ground state geometry optimization of the three forms of OxyLuc has been performed at VMC and DFT level. Semilocal (BLYP), hybrid (B3LYP), hybrid range-separated (CAM-B3LYP) and meta-GGA (M062X) functionals have been chosen for the DFT optimizations. The DFT structure calculations have been carried out using Gaussian 09 \cite{g09} with the diffuse-augmented polarized double-$\zeta$ basis set Def2-SVPD \cite{rap10}. \\
The VMC structural optimization has been obtained through the minimization of the energy with respect to both the variational parameters $ {\bm \alpha}$  and the nuclear coordinates $\textbf{R}$ of the trial wave function $\Psi_{T}$:

\begin{equation}
E^{OPT}_{VMC}=\min_{ {\bm \alpha}, \textbf{R}}  E\left[   \textbf{R} ;\Psi_T (\textbf{x}; {\bm \alpha}, \textbf{R}  )  \right].
\label{equ:E_VMC_R}
\end{equation}

The energy $E$ at a fixed configuration is evaluated as the expectation value of the electronic Hamiltonian $\hat{H}$, and the corresponding integral over the electronic coordinates $\textbf{x}$ (spatial $\textbf{r}$ and spin ${\bm \sigma}$) is written introducing the local energy $E_{L}(\textbf{x}; {\bm \alpha}, \textbf{R})=\hat{H}\Psi_{T}(\textbf{x}; {\bm \alpha}, \textbf{R})/\Psi_{T}(\textbf{x}; {\bm \alpha}, \textbf{R})$, and the probability density $ \Pi(\textbf{x}; {\bm \alpha}, \textbf{R}) =  \frac{\Psi_{T}^{2}(\textbf{x}; {\bm \alpha}, \textbf{R})}  {\int \Psi_{T}^{2}(\textbf{x}; {\bm \alpha}, \textbf{R}) d\textbf{x} } $:
\begin{equation}
 E[ \textbf{R} ;\Psi_T (\textbf{x}; {\bm \alpha}, \textbf{R}  ) ]= \int \Pi(\textbf{x}; {\bm \alpha}, \textbf{R}) E_{L}(\textbf{x}; {\bm \alpha}, \textbf{R}) d\textbf{x} = {\langle \hat{H} \rangle}_{\Pi}.
\label{eq:int1}
\end{equation}
The value of the integral in Eq. \ref{eq:int1} is estimated as a sum over a finite set of points in the $\textbf{x}$ space, generated by the Metropolis-Hasting algorithm according to the probability density $\Pi(\textbf{x}; {\bm \alpha}, \textbf{R}) $ \cite{metropolis}.\\
We have applied the linear method\cite{umr+07prl} with the inclusion of a partial Hessian to accelerate convergence\cite{sor+07jcp} for the optimization of the wave function variational parameters $\bm \alpha$, while the steepest descent algorithm has been used for the structure optimization.

Within the VMC framework, ionic forces for the nucleus $a$ are given by the following
expression: \cite{bk:hammond}

\begin{equation}
\textbf{F}_a (\textbf{R})=
-\left\langle \frac{d E_L}{d\textbf{R}_a} \right\rangle_{\Pi}
+2
\left\lbrace
\left\langle E_L\right\rangle_{\Pi}
\left\langle
\frac{d \ln \left[\Psi_T\right]}{d \textbf{R}_a}
\right\rangle_{\Pi}  \right.
  \left.-
\left\langle E_L
\frac{d \ln \left[\Psi_T\right]}{d \textbf{R}_a}
\right\rangle_{\Pi}
 \right\rbrace. 
\label{equ:total_force}
\end{equation}

The variance on the calculated forces is drastically reduced by using the Space Warp Coordinate Transformation (SWCT) \cite{Umrigar89,Assaraf03}.
Analytical derivatives of the function resulting by the combined use of pseudopotentials and SWCT, also taking into account the differentiation of the electronic coordinates with respect to the nuclear ones (beyond the expression reported in Eq. \ref{equ:total_force}), have been obtained using the adjoint algorithmic differentiation scheme \cite{sor+10jcp}. \\
VMC geometry optimization can be considered an accurate reference for the determination of structures of chromophores involved in biological processes \cite{Coccia:2012ex,Coccia14}, thanks to the explicit presence of the Jastrow factor \cite{zen14}. 
Another important advantage of the method is the extremely favorable parallelism of the algorithms, which is linear with the number of cores, allowing  to efficiently exploit high-performing computing facilities and, consequently, to massively apply QMC to the study of medium- and large-size chemical systems. \\
The trial wave function adopted in this work is the Jastrow Antisymmetrized Geminal Power (JAGP) \cite{cas+03jcp,cas+04jcp,zen14}, built as the product of an Antisymmetrized Geminal Power (AGP), defining the nodal surface, and a Jastrow factor. The latter is a positive function including many-particle terms for the correct description of the electron-electron and electron-nucleus cusp conditions\cite{dru+04prb} in the case of all-electron calculations, and of the dynamical correlation: 
\begin{equation}
\Psi_{T}(\textbf{x}) = \Psi_{AGP}(\textbf{x}) \times J(\textbf{r}).
\label{jagp}
\end{equation}
Note that the dependence of $\Psi_T$ from ${\bm \alpha}$ and $\textbf{R}$ will be omitted from here on.
For closed-shell molecular systems of $M$ atoms and $N_e$ electrons in a spin singlet state (as is the present case), where $N_e/2=N_e^\uparrow=N_e^\downarrow$, the AGP part is given by the antisymmetrized product:
%---------------
\begin{equation}
\Psi_{AGP}\left(\textbf{x}\right)=\hat{A}\prod_{i=1}^{N_{e}/2}\Phi_G \left( \textbf{x}^\uparrow_{i};\textbf{x}^\downarrow_{i} \right),
\label{equ:AGP}
\end{equation}
%---------------
where $\Phi_G$ are geminal functions of two electrons with opposite spin,
defined as the linear combination of products of two one-electron atomic orbitals centered on the different nuclei pairs. 
$\hat{A}$ is the antisymmetrization operator. \\
The Jastrow term $J$ is split  into a product of different terms $J = J_1 J_2 J_{3} J_{4} $ \cite{mar+09jcp, bar+12jctc,zen13}, accounting for one- ($J_1$), two- ($J_2$), three- ($J_3$) and four-body ($J_4$) contributions. The $J_3$ term refers to the $een$ contribution, while the $J_4$ one corresponds to $eenm$, with $n$ and $m$ being different nuclei. A detailed discussion on the role played by each contribution and on the functional form of the various Jastrow terms can be found in Refs. \cite{zen13} and \cite{zen14}. As shown in Eq. \ref{jagp}, the Jastrow factor is spin independent.

The TurboRVB package \cite{TurboRVB} has been used for the VMC calculations, following the computational protocol reported elsewhere \cite{Coccia:2012ex,Coccia14}. The basis sets used for the $\Psi_{AGP}$ and $J_3$ and $J_4$ terms of the trial wave function in the VMC geometry optimization are reported in Table \ref{tab:basis}. All linear and nonlinear parameters belonging to the AGP and the Jastrow terms have been optimized by minimizing the ground-state energy $E$. Pseudopotentials for C, N, O and S atoms have been employed \cite{Burkatzki:2007p25447,Burkatzki2008}. The effect of the pseudopotentials on the geometric parameters is negligible when compared with all-electron results, 
 as shown by previous calculations~\cite{bar+12jctc} for the C=C and C-H bonds of the ethylene molecule. 
\begin{table}
\begin{tabular} {c|c|c|c|c|c}
 & Carbon & Nitrogen & Oxygen &Sulfur & Hydrogen    \\ \hline
AGP  & (4s4p)/[2s2p] &  (4s4p)/[2s2p] & (4s4p)/[2s2p] & (4s4p)/[4s2p] & (3s1p)/[2s1p]  \\
\hline
$J_{3}$, $J_4$  &  (3s2p)/[2s1p]  &  (3s2p)/[2s1p] & (3s2p)/[2s1p]   & (3s2p)/[2s1p]   & (2s1p)/[1s1p] \\
\end{tabular}
\caption{\label{tab:basis} $\Psi_{AGP}$, $J_{3}$ and $J_4$ basis sets used for the VMC geometry optimization of keto-1, enol-1 and enol-2 forms of oxyluciferin.}
\end{table}

\subsection{Electronic and Optical properties}
Electronic and optical properties of the keto-1 and enol-1 forms have been calculated at DFT/TDDFT, CC2 and GW/BS levels. For enol-2, only DFT and TDDFT calculations have been performed.
Kohn-Sham frontier orbitals have been calculated using  Def2-SVPD, cc-pVTZ and aug-cc-pVTZ Dunning's basis sets.
TDDFT calculations have been carried out using BLYP, B3LYP, CAM-B3LYP and M062X functionals with Gaussian basis sets. For VMC and CAM-B3LYP structures we have employed basis sets up to the cc-pVQZ.\\
MBGFT excited-state calculations based on the GW/BS approach have been performed for the keto-1 and enol-1 forms. First we computed the single-particle Kohn-Sham states  $\phi_i^{KS}$ and corresponding energies $\epsilon^{KS}_i$, needed as starting-point in GW/BS procedure, then quasi-particle energies have been computed by considering the GW self energy in the quasi-particle equation. GW quasi-particle energies $\epsilon^{GW}_i$ are given by:
\begin{equation}
\epsilon^{GW}_i=\epsilon^{KS}_i+\langle \phi^{KS}_i \vert \Sigma^{GW}(\epsilon^{GW}_i)-V_{xc} \vert \phi^{KS}_i \rangle
\label{eqp}
\end{equation}
where $\Sigma^{GW}$ is the GW self energy, which is 
the product between the Kohn-Sham Green's function $G^{KS}$ and the screened Coulomb interaction $W$:
\begin{equation}
\Sigma^{GW}({\bf r, r^\prime},\omega) = \frac{i}{2\pi} \int d \omega^{\prime}  e^{i\eta  \omega^{\prime}} G^{KS}( {\bf r}, {\bf r}^{\prime},\omega+\omega^{\prime}) W({\bf r},{ \bf r}^\prime,\omega^\prime),
\label{sig}
\end{equation}
where the term $e^{i \eta \omega^\prime}$ enforces the correct time ordering of the self energy.
The screened potential $W$ is obtained within the random phase approximation (RPA) and $V_{xc}$ is the usual DFT exchange-correlation potential.

In order to remove the dependency of the final quasi-particle states on the specific  functional used for the DFT starting point, a partially self-consistent scheme (evGW) has been applied, where the $\epsilon_i^{GW}$ obtained from Eq.~\ref{eqp} are then reinserted in the construction of the Green's function and the polarizability in Eq.~\ref{sig} and Eq.~\ref{eqp} is iterated until self consistency is reached. In this procedure the Kohn-Sham wave functions are kept frozen. It has been shown by several authors that this partial self-consistent method strongly improves HOMO-LUMO gaps and thus BS excitation energies~\cite{jacquemin2015benchmarking,sharifzadeh2012quasiparticle,Baumeier2012,blase2011first,hogan2013correlation}. %In this procedure we have considered as starting point both eigenvalues and eigenfunctions coming from LDA and CAM-B3LYP approximation for the exchange and correlation functional, noticing that the evGW procedure strongly reduces the dependency from the initial condition for the case of keto-1 form, while for the enol-1 form such dependency is only softened (see below).
We noticed that, in the evGW procedure, the dependence of the results on the initial conditions is strongly reduced for the keto-1 form, while it is only slightly reduced for the enol-1 form (see below). Therefore, as starting eigenvalues and eigenfunctions, we used those obtained from the LDA and CAM-B3LYP approximations for the exchange and correlation functional.  
Excitation energies were then computed by solving the BS equation, i.e. including the electron-hole interaction (e-h). 
An electron-hole state can be described as:
\begin{equation}
\Phi^s({\bf r_e},{\bf r_h})=\sum_i^{occupied} \sum_a^{virtual}A^s_{ia}\phi_a({\bf r_e})\phi_j^*({\bf r_h})+B^s_{ia} \phi_j({\bf r_e})\phi_a^*({\bf r_h})
\end{equation}
where $A^s_{ia}$ and $B^s_{ia}$ are the resonant and anti-resonant amplitudes. 
In this basis the BS equation can be mapped onto a non-Hermitian eigenvalue problem~\cite{onida2002electronic,albrecht1998ab} and the $A^s_{ia}$ and $B^s_{ia}$ are then obtained as solutions of an excitonic Hamiltonian that reads: 
\begin{equation}
\begin{pmatrix}
R & C \\ -C^* & -R^*
\end{pmatrix}
\begin{pmatrix}
A^s\\
B^s
\end{pmatrix}
= E^{s}
\begin{pmatrix}
A^s\\
B^s
\end{pmatrix}
\label{bsematrix}
\end{equation}
In the excitonic Hamiltonian
$R$ is the Hermitian resonant part:
\begin{equation}
R=(E^{GW}_a-E^{GW}_i)\delta_{i,j}\delta_{a,b}+\langle ai \vert K \vert jb\rangle
\end{equation}
and $C$ is the coupling symmetric part:
\begin{equation}
C=\langle ai \vert K \vert \overline{ jb }\rangle,
\end{equation}
where $K = W - 2V$ is the BS excitonic kernel, with $W$ and $V$ the screened and bare Coulomb interaction respectively, and  $\overline{jb}$ indicates the electron-hole anti-pairs.
MBGFT calculations have been performed as implemented in the all-electron Gaussian basis set MolGW~\cite{bruneval2016molgw} package using the cc-pvQZ  basis set and resolution-of-the-identity (RI) technique~\cite{weigend2002fully}. The RI technique expresses four-center integrals in terms of two and three-center integrals by using auxiliary basis sets.
All virtual states are included in the construction of both the polarizability and the self-energy,  
while core states are neglected. Quasi-particle energies of Eq.~\ref{eqp} have been computed via graphical solution without resorting to linearization around Kohn-Sham energy for the correlation part of the self-energy. The BS equation has been solved by considering the full matrix of Eq.~\ref{bsematrix}, beyond the Tamm-Dancoff approximation. For the keto-1 form we have also calculated excitations at evGW/BS level using the plane-wave code Yambo~\cite{marini2009yambo} starting from the ground-state electronic structure calculated by the Quantum ESPRESSO package~\cite{giannozzi2009quantum}, obtained by
Troullier-Martins norm-conserving pseudo-potentials  with a plane-wave cutoff of 60 Ry for the wave functions.
The spurious interactions between images coming from periodic boundary conditions have been avoided by truncating the long-range Coulomb interaction using the cutoff technique described in Ref.~\cite{rozzi2006exact}.
For the GW correction 1000 states have been included to calculate the dynamical dielectric matrix  (8 Ha cutoff) and the Green's function. 
The dynamical screening is treated within the plasmon-pole approximation~\cite{godby1989metal}.
For the BS spectra, we have included electron-hole pairs considering 33 occupied and 95 unoccupied states. 1500 states have instead been included in the static polarization function. We have used a cutoff of 10 Ha for the screened interaction and of 20 Ha for the exchange component. 

The Molpro package \cite{MOLPRO} has been used for the CC2 excited-state calculations with the cc-pVDZ and cc-pVTZ Dunning's basis set, while, unfortunately, cc-pVQZ basis set produces memory and instability issues. 

We have also performed DFT geometry optimizations and TDDFT excited-state calculations in implicit solvent with the polarizable continuum model (PCM) \cite{tom05} using the Gaussian 09 code. In detail,
keto-1, enol-1 and enol-2 forms have been optimized in water and toluene using the Def2-SVPD basis set at DFT/CAM-B3LYP level of theory. The same functional has been used for the TDDFT calculations. In order to assess the stability of the dianionic  enol-2 in gas phase, frontier orbitals and HOMO-LUMO gap 
have been computed with BLYP, B3LYP, CAM-B3LYP and M062X functionals and cc-pVTZ, aug-cc-pVTZ and Def2-SVPD basis sets. 

\section{Results and discussion}
\label{res}
Starting from the consensus about the possible role as main emitters of keto-1, enol-1 and enol-2 forms, we report our results on the structural properties in Section \ref{structure}. 
The vertical S$_1 \leftarrow $ S$_0$ excitations for keto-1 and enol-1 are presented in Section \ref{exc}, and in Section \ref{enol2} we discuss the excitations of all three forms in the presence of a solvent, emphasizing the intrinsic difficulty to treat the enol-2 dianion in the gas phase.

\subsection{S$_0$ gas phase structures}
\label{structure}
In this section we compare a selection of bond lengths obtained relaxing the ground-state structures at different levels of theory. These geometric parameters are key quantities to assess the quality of the ground state and to properly compute the vertical absorption energies. The comparison aims at assessing the role of the computational method in determining geometric parameters for a given form, and to show the differences in the geometrical parameters among the three forms, when the structures are optimized using the same approach. 

To simplify the discussion, we have numbered the involved atoms as reported in Figure \ref{fig1}. In detail, we have focused the attention on the C$_{2'}$-C$_2$ central bridge bond, which is characterized by an hybrid nature of single and double bond, the C$_2$-N$_3$ and N$_{3}$-C$_{4}$ bonds, the C$_5$-S$_1$ bond and the C$_4$-C$_5$ bond on the five-ring moiety, together with the two C-O bonds. All the values are reported in Table \ref{tab2}. The bond lengths between two atoms $m$ and $n$ will be indicated as R$_{m,n}$. 
Following the analysis reported in Ref. \cite{Cheng15}, we preferred to analyze those bonds that display a resonance character, and which are therefore involved in the $\pi$ conjugation of the OxyLuc forms. In Table S1 of the Supporting Information (SI) we also report the 3-2-2' and 1-2-2' angles and the 2-2-2'-3' dihedral for the optimized structures of the three forms. 

Starting from the keto-1 form, we observe that the CASSCF(18,15) \cite{liu08} bond length for C$_{2'}$-C$_2$ is the shortest (1.401 \AA), corresponding to a (slightly) smaller flexibility with respect to the structures obtained with the other methods. This is a quite expected result, since it is well known that a proper description of the dynamic correlation (reducing the difference between single and double bonds in a conjugated moiety) is lacking in the CASSCF representation of the wave function. On the other hand, VMC (1.408(3) \AA), CAM-B3LYP (1.412 \AA), M062X (1.415 \AA) and B3LYP \cite{sto13} (1.414 \AA) values are very close to each other, with the exception of BLYP (1.420 \AA). 

For what concerns the two C-N bonds in the (formally) -N=C-C=N- moiety, a small but significant asymmetry in the bond length is found, with R$_{2',3'}$ longer than R$_{2,3}$  in all the structures considered in this work. Interestingly, the difference between R$_{2',3'}$ and R$_{2,3}$ increases when reducing the amount of dynamic correlation, moving from BLYP to CASSCF(18,15): the calculated differences are 9 m\AA~for BLYP, 12 m\AA~for B3LYP, 14 m\AA~for CAM-B3LYP and M062X, 16(3) m\AA~for VMC and 25 m\AA~for CASSCF(18,15). CAM-B3LYP calculations produce bond distances in very good agreement with the VMC findings.  

Besides the CASSCF(18,15) results, the VMC geometry shows the shortest double bond R$_{4,5}$, 1.538(2) \AA. This finding is in fair agreement with what reported in the previous study of the VMC/JAGP computational protocol on linear conjugated chromophores as the retinal protonated Schiff base \cite{Coccia:2012ex}, the peridinin carotenoid \cite{Coccia14} and polyacetilene fragments \cite{bar+15jctc_b}: BLYP and B3LYP functionals overcorrelate  the polyenic chain, resulting in a small bond alternation length (i.e., too long double carbon bonds, as in this case), while VMC/JAGP produces a more balanced description of the electronic $\pi$-delocalization.

The VMC estimate of R$_{5,1}$, involving the sulfur atom, gives the shortest bond length (1.796(1) \AA), around 10 m\AA~smaller than the CAM-B3LYP value (1.807 \AA). 
R$_{4,11'}$ and R$_{6',10'}$ oxygen-carbon bonds are not dramatically affected by the method chosen for the geometry optimization, with CAM-B3LYP (and M062X) and VMC producing very similar bond lengths; the only exceptions are given by the CASSCF(18,15) and BLYP, with respectively shorter and longer bond lengths. For R$_{3,4}$, on the other hand, all the methods show bond distances in a range of 10 m\AA.  

An accurate description of the bond length pattern is essential for the computation of vertical absorption energies in biochromophores \cite{Coccia:2012ex,Coccia14,var14}. Therefore a comparison of the obtained structures with the one obtained with highly accurate quantum chemistry methods as CCSD(T) would be highly desirable. To the best of our knowledge, unfortunately, no CSSD(T) result is present in the literature on geometry optimization of (nonsymmetric) biological chromophores. VMC structures on polyenic chains have been shown to be as accurate as the CCSD(T) ones for the bond length alternation \cite{bar+15jctc_b}, therefore we are confident that the VMC structure can be taken as the theoretical reference for assessing the quality of various DFT approaches. Here, we clearly see that the CAM-B3LYP functional provides structures in quantitative agreement with the VMC ones, with a maximum discrepancy of 11 m\AA~, thus enforcing the idea that the use of CAM-B3LYP for the geometry optimization of chromophores is a fully reliable choice as already observed in previous studies \cite{Coccia:2012ex,Coccia14}. This conclusion can be also extended to M062X. 

Qualitatively, the various methods adopted for the geometry optimization of keto-1 behave similarly when applied to enol-1 and enol-2. 
Looking at the specific features of enol-1 and enol-2 forms,
in the VMC enol-1 structure we found a bridge bond R$_{2,2'}$ of 1.434(1) \AA, longer than in keto-1 (1.408(3) \AA). 
The large differences  found for  R$_{4,5}$ and R$_{4',11'}$  between keto-1 and enol-1 are obviously due to the keto-enol tautomerization. A rather large difference is also seen for the bond involving the sulfur atom, with a shortening of 56(2) m\AA. 
The C$_{2'}$-C$_{2}$ is even longer in the enol-2 form (1.457 \AA). 

Generally, the comparative analysis of keto-1, enol-1 and enol-2 forms allows us to state that the level of theory used for the geometry optimization of the three species does not affect much the relative structural properties.

\begin{table}
\scriptsize
\begin{tabular} {c|c|c|c|c|c|c|c|c}
 & R$_{2',2}$ & R$_{2,3}$ & R$_{2',3'}$ & R$_{3,4}$ & R$_{4,5}$ &R$_{5,1}$ & R$_{4,11'}$ &  R$_{6',10'}$    \\ \hline
 \\[\dimexpr-\normalbaselineskip+3pt]
{\bf keto-1} \\[.1cm]
BLYP & 1.420 & 1.326 & 1.335 & 1.380 & 1.565  & 1.841 & 1.235  &  1.267  \\
 B3LYP$^{a}$ & 1.414 & 1.313 & 1.325 & 1.371 & 1.550  & 1.821 & 1.221  &  1.251   \\
 M062X & 1.415 & 1.304 & 1.318 & 1.375 & 1.545 & 1.806 & 1.210 &  1.241 \\
 CAM-B3LYP & 1.412 & 1.305 & 1.319 & 1.370  & 1.541  & 1.807  &1.214  &  1.245  \\
VMC & 1.408(3) &  1.307(2) & 1.323(2) & 1.376(1) & 1.538(2) & 1.796(1) & 1.211(1)  &   1.243(3)  \\
CASSCF(18,15)$^{b}$&  1.401 & 1.292  & 1.317  & 1.375 & 1.523 & 1.803 & 1.203  &  1.230  \\
\hline 
\\[\dimexpr-\normalbaselineskip+3pt]
{\bf enol-1}\\[.1cm]
 BLYP & 1.432 & 1.335 & 1.325 & 1.365 & 1.382  & 1.767 & 1.378  &  1.271  \\
 B3LYP & 1.430 & 1.321 & 1.312 & 1.356  & 1.370 & 1.750 & 1.361 &  1.257   \\
 M062X & 1.437 & 1.311  & 1.302 & 1.357  & 1.365   & 1.736  & 1.354  &  1.248   \\
 CAM-B3LYP & 1.436 & 1.311 & 1.302  & 1.356  & 1.362   & 1.739  & 1.354 &  1.252   \\
 VMC & 1.434(1) & 1.312(2) & 1.303(1) & 1.361(2) & 1.354(1)  &  1.730(2) & 1.357(2) &  1.252(1)  \\
 CASSCF(18,15)$^{b}$ & 1.436 & 1.294  & 1.292  & 1.361 & 1.349 & 1.738 & 1.350  &  1.238  \\
\hline
\\[\dimexpr-\normalbaselineskip+3pt]
{\bf enol-2}\\[.1cm] 
BLYP & 1.452 & 1.315 & 1.316 & 1.434 & 1.434  & 1.760 & 1.273  &  1.288  \\
B3LYP & 1.452 & 1.302  & 1.301  & 1.422  & 1.420 & 1.745 & 1.261  &  1.273   \\
 M062X & 1.462 & 1.292 &  1.290 & 1.424 & 1.413  & 1.735 & 1.254 & 1.264   \\
CAM-B3LYP & 1.460 &  1.293 &  1.290 & 1.421  & 1.409  & 1.738 & 1.257  & 1.269   \\
 VMC & 1.457(2) & 1.293(2) & 1.292(2) & 1.430(3)  & 1.404(4)  & 1.728(4) & 1.258(2)  &  1.268(1)   \\
 CASSCF(18,15)$^{b}$ & 1.460  & 1.278  &  1.281  &  1.434 & 1.383 & 1.752 & 1.246 &1.253  \\

\end{tabular}
\caption{\label{tab2} Optimized bond lengths (in \AA) for the VMC and DFT (BLYP, B3LYP, M062X and CAM-B3LYP) structures of keto-1, enol-1 and enol-2. Def2-SVPD basis set has been used for DFT calculations. $^{a}$Geometry from Ref. \citenum{sto13}. $^{b}$CASSCF(18,15) structures, with the the ANO-RCC-VDZP basis set, from Ref. \citenum{liu08}.}
\end{table}

Trans isomers with respect to the bridge R$_{2'-2}$ bond have been reported to be more stable than the cis ones in gas phase, as shown by simple calculations using the Boltzmann distribution and molecular dynamics simulations \cite{sto13}. 
At room temperature, the isomerization from trans to cis of the keto-1 form has substantially zero probability to occur. 

Moreover, at MS-CASPT2\cite{chen11} and DFT\cite{sto13} levels of theory, the keto-1 form is significantly more stable than the enol-1, with trans-cis energy differences of 13.4 and 11.3 kcal/mol, respectively. Thus, assuming that in the gas-phase experiment only the keto-1 form exists, we are allowed to compute the isomerization energy of the keto-1 species. On the other hand, the enol-1 form was found to be slightly more stable than keto-1 in water clusters \cite{nog16}.
Using the Def2-SVPD basis set we have estimated the trans-cis energy difference at DFT level for the keto-1 form, obtaining 12.7 kcal/mol with BLYP, 9.5 kcal/mol with CAM-B3LYP and 7.3 kcal/mol with M062X. 
The CAM-B3LYP energy  difference (taken as a representative example) diminishes when the two forms are embedded in a solvent: 7.1 kcal/mol in water and 8.2 kcal/mol in toluene.    
These findings confirm the fact that the experimental absorption energy corresponds to the trans isomer of the keto-1 form: indeed, the fraction of the enol-1 form, even using the smallest energy gap, is less than $10^{-5}$. 

In Ref. \cite{sto13}, the tautomer identity of the OxyLuc moiety in gas phase was ascertained by the absorption experiment done using the 5,5-dimethylated OxyLuc anion, where the molecule is locked as the keto-1 form by substitution. The red-shift observed in the absorption maximum matches well with the substitution effects measured in solution and computed at EOM-CCSD and TDDFT level. 

The results of this analysis, of course, can not easily be extended to OxyLuc embedded in the various protein environments, where, as explained in the introduction, many factors occur in defining the role of the emitter species. 

Even though literature \cite{chen11,sto13} and present results agree in identifying the trans isomer of the keto-1 form as the most stable moiety in gas phase at ambient conditions, the systematic comparison carried out between the bond lengths of keto-1, enol-1 and enol-2 represents an useful investigation on the correlation between geometric and optical properties, as reported in the next Subsection.

\subsection{S$_1 \leftarrow $ S$_0$ vertical absorption for keto-1 and enol-1}
\label{exc}
In gas-phase the only existing OxyLuc form is keto-1, as inferred from the stability arguments reported above. From a theoretical standpoint, therefore, we now wish to compare the experimental absorption of 2.26(8) eV (548(10) nm) obtained using the action spectroscopy technique \cite{sto13} with the vertical excitation energy computed on the keto-1 form. In order to achieve this goal, we computed the S$_1 \leftarrow $ S$_0$ excitation energy ($\Delta$E) using TDDFT with several functionals, evGW/BS (with Gaussian basis sets and plane waves), and CC2 on structures optimized at different levels of theory (VMC, DFT and CASSCF). Large variations exist in the $\Delta$E value according to the chosen computational protocol. Moreover, the same systematic study on the optical properties of enol-1 revealed that the absorption energy shift of enol-1 with respect keto-1 can change sign when the theoretical approach and the convergence degree with respect to the basis set are changed.

\begin{figure}[tbp]
\includegraphics[width=0.75\columnwidth]{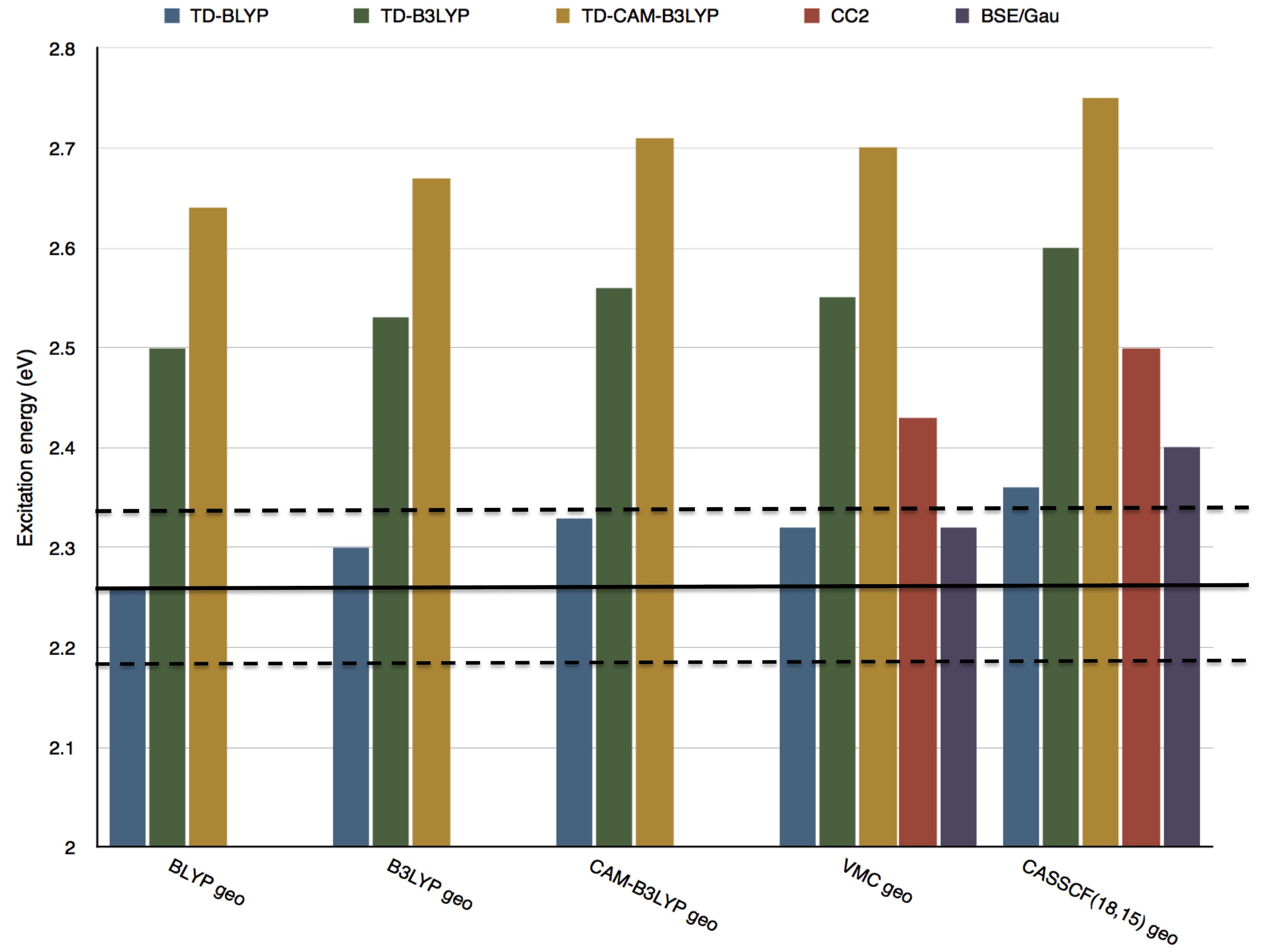} \\
\caption{Excitation energies (in eV) for the keto-1 form using TDDFT (BLYP, B3LYP\cite{sto13} and CAM-B3LYP with Def2-SVPD basis set), CC2 (cc-pVTZ basis set) and evGW/BS (cc-pVQZ basis set).  Optimized ground-state geometries from DFT (BLYP, B3LYP and CAM-B3LYP), CASSCF(18,15)\cite{liu08} and VMC. Horizontal lines correspond to the experimental value of 2.26 eV (solid) and the related error ($\pm$ 0.08 eV, dashed). The corresponding TDDFT values are collected in Table S2 of SI. \label{fig2} }
\end{figure}

The S$_1 \leftarrow $ S$_0$ transition studied in this work is characterized by a predominant single-excitation nature \cite{sto13}.
TDDFT excitation energies calculated on the different structures of keto-1 form obtained as described above are shown in Figure \ref{fig2}. It is evident that TDDFT results show a poor agreement with the experimental value, with the exception of 
the semi-local BLYP functional, probably because of a cancellation of errors, as already suggested in Ref. \cite{sto13}.
 TDDFT calculations have been performed using the Def2-SVPD basis set, as done in Ref. \cite{sto13} and, as pointed out in the same reference, 
the results are clearly affected by the choice of the exchange correlation kernel. 
On the other hand, it is important to stress that the fine convergence with respect to the basis set is of fundamental importance when determining the relative shift in energy between the keto-1 and enol-1 excitations. The complete list of results is reported in SI in Tab. S2 and S3.

\begin{figure}[tbp]
\includegraphics[width=0.75\columnwidth]{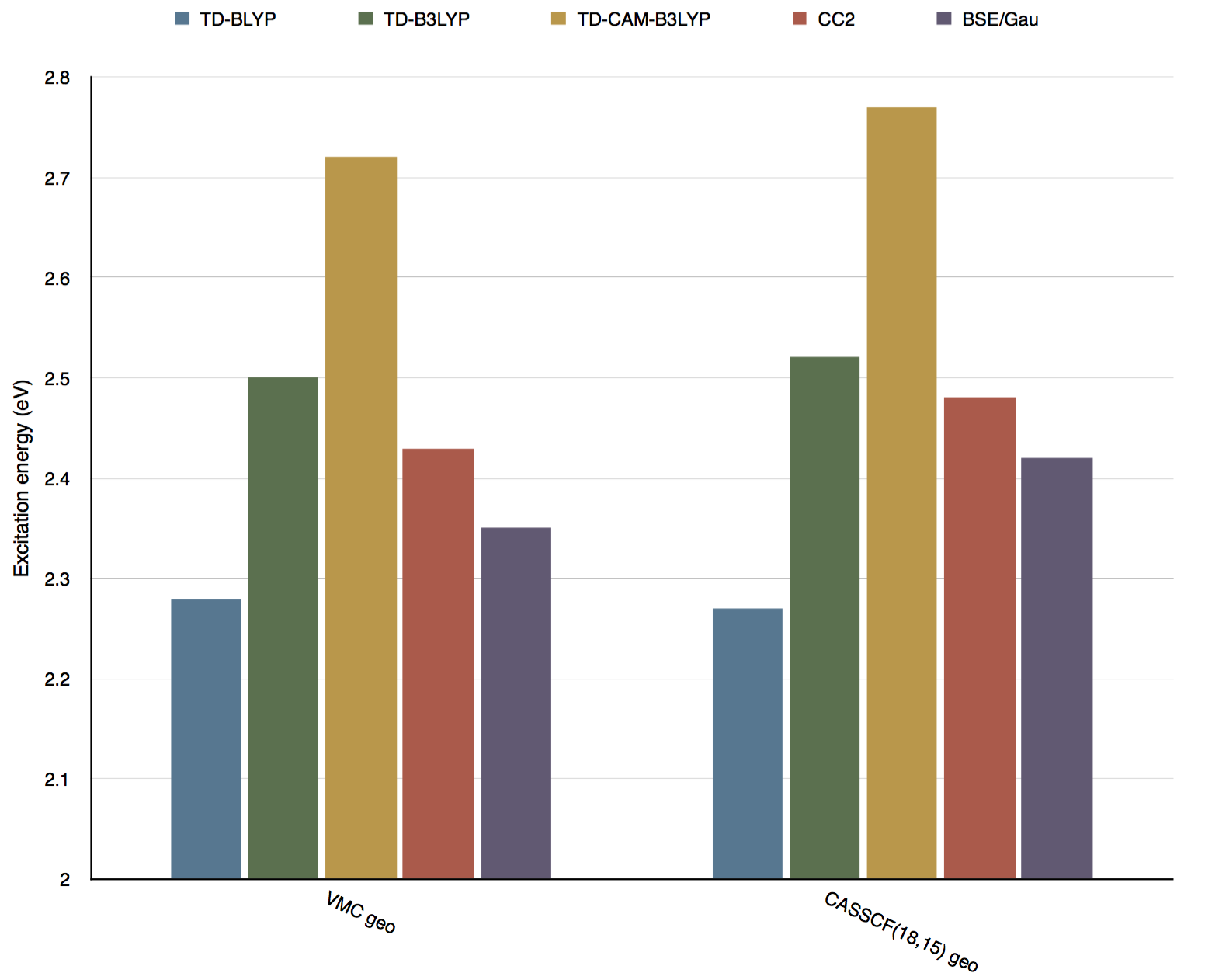} \\
\caption{
Excitation energies (in eV) for the enol-1 form using TDDFT (BLYP, B3LYP and CAM-B3LYP with Def2-SVPD basis set), CC2 (cc-pVTZ basis set) and evGW/BS  (cc-pVQZ basis set).  Optimized ground-state geometries from VMC and   CASSCF(18,15)\cite{liu08}. The corresponding TDDFT values are collected in Table S2 in SI. \label{fig3}}
\end{figure}

From Fig.~\ref{fig2} it is clear that B3LYP and CAM-B3LYP functionals overestimate the excitation energy of the keto-1 form, regardless of the optimized structure: the B3LYP excitation energy spans from 2.50 to 2.60 eV, the CAM-B3LYP is always larger than 2.60 eV. TDDFT values with the M062X functional show the same trend of those using CAM-B3LYP (Table S2 in SI). As expected, the net effect of the ground-state structure is to induce a blue shift, when the same excited-state approach is employed, in conjunction with a decrease of the correlation along the conjugated bonds: the CASSCF(18,15) structure produces the highest S$_1$ energy, as also found for the enol-1 form as shown in Figure \ref{fig3}.  In summary, the level of theory employed for the structural optimization produces a moderate blue shift in the vertical excitation, smaller than 0.1 eV when the difference between single and double bonds increases. This is a much more reduced effect than those observed for linear chromophores like the peridinin carotenoid \cite{Coccia14}. We note here that CAM-B3LYP and VMC geometries determine essentially indistinguishable excitations for the keto-1 form (Figure \ref{fig2} and Table S2), provided the same functional for TDDFT is used.

TDDFT excitation energies computed using the 6-311+G* basis set,  reported in Table S3 of SI, show no substantial difference with the Def2-SVPD set of results. 

Excluding the poor TD-BLYP results, only the combination of VMC and BS is able to accurately match with the experimental value of 2.26(8) eV, as shown in Figure \ref{fig2}. The evGW/BS excitation energy for the keto-1 form (2.32 eV) was obtained using the cc-pVQZ Gaussian basis set starting from CAM-B3LYP eigenvalues and eigenstates.
We have tested the dependence on the choice of the exchange and correlation functional in the ground state in the partially self-consistent evGW/BSE procedure using also LDA functional obtaining a very similar result, 2.34 eV (see Table S5).  
The convergence study (on the VMC structure of the keto-1 form) of the excitation energy as a function of the size of the basis set is reported for TDDFT (CAM-B3LYP) and evGW/BS in Tables S4 and S5 of SI, respectively: the energy decreases when increasing the complexity of the basis set, the cc-pVQZ being the largest basis set one can reasonably apply to OxyLuc systems. The convergence for the evGW/BS excitation energy has been also validated with respect a plane wave calculation starting from LDA ground state (Table S5).

Similarly, we report in Figure \ref{fig3} the values of the excitation energy for the enol-1 form, using the ground-state optimized VMC and CASSCF(18,15) structures, at TDDFT (BLYP, B3LYP and CAM-B3LYP), CC2 and evGW/BS level. TDDFT excitation energies are also reported in S2 of SI, while BS and CC2 energies are in Tables \ref{bse} and \ref{cc2}, respectively. The CASSCF(18,15) structure produces a systematic shift towards larger energies regardless of the excited-state approach. In particular, the net effect of using the CASSCF(18,15) geometry of the enol-1 form is to increase the excitation energy by 0.07 eV, when calculated in GW/BS framework, and by 0.05 eV for CC2.
Also in this case, the TDDFT results span a large energy range (around 0.4 eV) according to the functional employed for the calculation, regardless of the chosen geometry.

Looking at the keto-1/enol-1 S$_1$ energy shift (defined as the difference between the keto-1 and the enol-1 excitation energies, see Table \ref{gap}), one can compare the behavior of the different excited-state techniques applied on the VMC and CASSCF(18,15) geometries. A robust computational reference for the gas phase S$_1 \leftarrow$ S$_0$ excitation energies is given by the MS-CASPT2 results \cite{chen11}, even though the transition energies slightly overestimate the experimental transition value (2.43 eV for the keto-1 form): the MS-CASPT2 keto-1/enol-1 shift (computed on the CASSCF(18,15) structures) is $-$0.05 eV. 
The relative keto-1/enol-1 shifts computed at different levels of theory are collected in Table.\ref{gap}.
At TDDFT level, only the use of CAM-B3LYP and M062X functionals on the VMC structures can correctly reproduce the sign and the magnitude of the shifts predicted by the accurate MS-CASPT2 analysis. Indeed, a shift of $-$0.02 (CAM-B3LYP) and $-$0.03 eV (M062X) is found for the enol-1 form with respect to the keto-1 one, while BLYP and B3LYP yield a positive S$_1$ energy gap.
The BS shift is characterized by the same sign of the MS-CASPT2 one, and an absolute value equal to that of TDDFT with CAM-B3LYP and M062X: $-$0.03 (VMC geometry) and $-$0.02 eV (CASSCF(18,15) geometry).

 It is important to note that while the keto-1 evGW/BS excitation energy is only barely affected by the choice of the DFT initial guess, passing from CAM-B3LYP (2.32 eV) to LDA (2.34 eV), a large effect (i.e. blue shift) is instead observed for enol-1 passing from 2.35 eV to 2.45 eV when LDA ground state is considered as a starting point. Such a difference needs to be ascribed to the different localization of the wave functions at CAM-B3LYP and LDA level for the enol-1 form. The LDA-based partially self consistent evGW/BS excitations for VMC and CASSCF(18,15) structures are reported in Tab. S7. 

Using a cc-pVTZ basis set, the CC2 shift is zero when the VMC structure is used, and 0.02 eV for the CASSCF(18,15). Unfortunately, CC2 calculations with a cc-pVQZ basis set are not affordable since they are extremely memory demanding; however, the trend in the excitation energies (see Table S6) 
passing from cc-pVDZ to cc-pVTZ indicates that, for an even larger basis set, the sign of the keto-1/enol-1 shift could be the same of the one obtained in evGW/BS and MS-CASPT2 framework. \\
%To conclude, a small red-shift has been observed in the absorption of the cis isomer of the keto-1 form with respect to the trans one. Results using TDDFT (BLYP, B3LYP, CAM-B3LYP and M062X) on the VMC structure are collected in Table S9 in Supporting Information. With the same Def2-SVPD basis set, the red-shift is 0.07 eV when adopting the BLYP functional, and only 0.01 eV with CAM-B3LYP and M062X.

\begin{table}
\begin{tabular} {cc|c|c}
 S$_{0}$ Geo  & Form & $\Delta$E (nm) & $\Delta$E (eV)    \\ \hline

\multirow{2}{*}{VMC} & keto-1  & 534 &  2.32   \\
& enol-1   & 528 &  2.35  \\
\hline 
\multirow{2}{*}{CASSCF(18,15)$^a$} & 
keto-1  & 517 &  2.40   \\
& enol-1  & 512 & 2.42   \\ 

\end{tabular}
\caption{\label{bse} evGW/BS vertical S$_1 \leftarrow$ S$_0$ absorption energies using the cc-pVQZ basis set for the keto-1 and enol-1 forms with VMC and CASSCF structures. CAM-B3LYP ground state is used as starting point for the partially self-consistent evGW.
The experimental excitation energy is 2.26 $\pm$ 0.08 eV (548 $\pm$ 10 nm). $^a$Structure from Ref. \citenum{liu08}.
}
\end{table}

\begin{table}
\begin{tabular} {cc|c|c}
 S$_{0}$ Geo  & Form &  $\Delta$E (nm) & $\Delta$E (eV)    \\ \hline
\multirow{2}{*}{VMC} & keto-1  &  510  & 2.43       \\  
& enol-1 &  510  &  2.43          \\ \hline  
\multirow{2}{*}{CASSCF(18,15)$^a$} & 
keto-1   & 496  & 2.50       \\ 
& enol-1 & 500  & 2.48         \\   
\end{tabular}
\caption{\label{cc2}  CC2 vertical vertical S$_1 \leftarrow$ S$_0$ absorption energies using the cc-pVTZ basis sets on keto-1 and enol-1 forms with VMC and CASSCF structures.  $^a$Structure from Ref. \citenum{liu08}.
The experimental excitation energy is 2.26 $\pm$ 0.08 eV (548 $\pm$ 10 nm).
}
\end{table}

\begin{table}
\begin{tabular} {c|c|c|c}
 S$_{0}$ Geo & S$_{1} \leftarrow$ S$_{0}$ & Shift
 (nm) & Shift (eV)    \\ \hline
\multirow{6}{*}{VMC} & BLYP & -9 & 0.04    \\
& B3LYP & -10 & 0.05    \\
& CAM-B3LYP & 4 & -0.02 \\
& CAM-B3LYP$^a$ & 3 & -0.02 \\
& GW/BS  & 6   & -0.03    \\
& CC2 & 0    &  0.00      \\ \hline
\multirow{6}{*}{ CASSCF(18,15)$^{a}$} & MS-CASPT2$^{b}$ &  10 & -0.05     \\ 
& BLYP & -23  &  0.09    \\ 
& B3LYP & -15  & 0.08     \\ 
& CAM-B3LYP & 3  & -0.02     \\
& GW/BS  & 5  & -0.02     \\
& CC2 & -4  & 0.02     \\ 
\end{tabular}
\caption{\label{gap} Relative shifts between keto-1 and enol-1 forms  for vertical absorption energies. The Def2-SVPD basis set has been used for the TDDFT calculations. $^a$Structures from Ref. \citenum{liu08}. $^{b}$Data from Ref. \citenum{chen11}.} 
\end{table}

\subsection{S$_1 \leftarrow$ S$_0$ vertical absorption for enol-2}
\label{enol2}
The study of the ground and excited-state properties of the gas phase enol-2 form represents a challenge for any theoretical approach, since a double negative charge characterizes the electronic structure of the system (Figure \ref{fig1}). A deeper insight into the properties of the HOMO and LUMO energies is therefore needed in order to understand whether the two electrons in excess can maintain the electronic density bound in gas phase conditions, without a stabilizing counterion. 

The HOMO and LUMO energies of the gas-phase keto-1, enol-1 and enol-2 forms are reported in Table S8 of SI. These calculations were done considering the CAM-B3LYP structures, which, as previously mentioned, are close to the VMC ones.
First we observe that for the singly-charged systems, the HOMO orbital is bound with a negative energy, regardless of the specific choice for the calculation, whereas the LUMO orbital lies in the continuous part of the eigenvalue spectrum. Energies are sensitive  to the addition of diffuse functions to the cc-pVTZ basis set, but the gap remains substantially unchanged (the largest difference being of 0.1 eV for the enol-1 using M062X and aug-cc-pVTZ), confirming that the electronic density is well localized around the nuclei of the keto-1 and enol-1 species. \\
The behaviour of HOMO and LUMO energies, and of the corresponding gap dramatically changes when looking at the enol-2 form, showing a strong dependence on the chosen basis set. The HOMO orbital is characterized by a positive energy, which significantly decreases (about 0.3-0.4 eV) when augmentation is added. A similar enhanced effect is found for the LUMO energy, decreasing of about 1.0-1.8 eV when passing from cc-pVTZ to aug-cc-pVTZ, thus displaying an evident decrease of the gap.  
The same issue regarding the convergence of the gap is found when plane waves are used as basis. In Table S9 of SI, DFT and Hartree-Fock HOMO and LUMO energies for increasing supercell size are reported.  DFT HOMO energy is always positive regardless the box size, while the LUMO energy drastically reduces when the simulation box increases. At Hartree-Fock level, the HOMO orbital becomes weakly bound when increasing the box size, while a not clear convergence  is seen for LUMO; moreover, convergence issues in the DFT self consistent cycle arise when even larger simulation boxes are used.

The results of the tests performed both in Gaussian basis set and in plane waves clearly show that ground state density for the gas phase enol-2 is unbound, thus questioning the reliability of previous gas phase calculations present in the literature.

The ANO-RCC-VDZP basis set used to compute MS-CASPT2 absorption of the gas phase enol-2 in Ref. \cite{chen11} contains diffuse functions, 
but this does not guarantee the robustness of the results for the reason above.
Convergence studies with respect to the basis set size for the enol-2 form have then been carried out including solvation effects via
 a polarizable continuum  model \cite{tom05} considering the chromophore embedded in water and toluene. 
Consistently with the calculations shown above, geometry relaxation of the three forms in solvent has been performed using the CAM-B3LYP functional and the Def2-SVPD basis set.
The values of the HOMO-LUMO gaps obtained with three functionals for three basis sets are reported in Table \ref{tab:solv}. The corresponding energies are instead given in Table S10 of SI. HOMO and LUMO energies and the corresponding gaps for the keto-1 and enol-1 forms in water and toluene are instead reported in Tables S11 and S12, respectively.
The direct interaction with a polarizable medium stabilizes the electronic density of the dianion, making the HOMO-LUMO gap less sensitive to the basis set, while, as expected, an evident dependence on the DFT functional is still present. In water, both HOMO and LUMO energies become negative and the corresponding gap (for a given functional) is not strongly affected by the augmentation. The same stability is found for the enol-2 in toluene, a non-polar solvent, even if only the HOMO orbital has a bound energy.

\begin{table}[!t]
\footnotesize
\tabcolsep=0.01\textwidth
\caption{HOMO-LUMO gaps (eV) at DFT level for enol-2 in water and toluene, using different basis sets. VTZ and aVTZ are abbreviations for cc-pvTZ and aug-cc-pvTZ, respectively. }
\label{tab:solv}
\begin{tabular}{c|ccc|ccc|ccc}
         & \multicolumn{3}{c}{BLYP}  &  \multicolumn{3}{c}{B3LYP} &  \multicolumn{3}{c}{CAM-B3LYP} \\ \hline 
eV & VTZ    &   aVTZ   & Def2-SVPD &     VTZ           &  aVTZ     & Def2-SVPD &    VTZ & aVTZ & Def2-SVPD \\ \hline
 {\bf water} & 1.70  &   1.72   & 1.73 &  2.90 &   2.91 & 2.92 &      5.24  &  5.23 &  5.25 \\ 
{\bf toluene}& 1.65 &  1.67  & 1.68 &   2.85 &  2.85 & 2.88 &  5.20&  5.18  & 5.21 \\
\end{tabular}
\end{table}

The S$_1 \leftarrow$ S$_0$ transition on the CAM-B3LYP structures of the three forms in water and toluene have been calculated using the CAM-B3LYP and the Def2-SVPD basis set. The results are shown in Table \ref{tab:solvent}, together with the gas phase value for keto-1 (see Table S2) and enol-1.

It is interesting to note that the three forms are characterized by a different response with respect to the solvent: keto-1 excitation energy is red-shifted  with respect to the gas phase, with only a small variation in water (0.04 eV); a large blue-shift is seen for the enol-1 form in water, while only small differences are found in toluene.

In water, enol-1 shows the maximum excitation energy; our calculations qualitatively confirm the relative shift in water among the three forms already computed at TDDFT/CAM-B3LYP level using the 6-31G$^{**}$ basis set \cite{Cheng15}, though the $\Delta E$ for keto-1 and enol-1 are sensibly larger (2.89 and 3.23 eV, compared with our values 2.67 and 3.00 eV), indicating a not negligible basis and geometry effect. 
On the other hand, our calculations provide the same excitation energies in water of the one found in Ref.~\cite{hiy13} at 
PCM/aug-cc-pVTZ level. More significantly, our results show a quantitative agreement with the experimental findings of Ref. \cite{gho15} for the three forms in water.

\begin{table}
\begin{tabular} {cc|c|c|c}
Form & Gas phase &  Toluene & Water & Exp in Water$^a$    \\ \hline
keto-1 & 2.71 & 2.53 & 2.67 & 2.56    \\
enol-1 & 2.73 & 2.71 & 3.00 & 3.05   \\
enol-2 & nc & 2.85 & 2.94 & 2.92\\

\end{tabular}
\caption{\label{tab:solvent} TDDFT/CAM-B3LYP absorption energies (in eV) for the three OxyLuc forms using the Def2-SVPD basis set and CAM-B3LYP structures optimized in gas phase, water and toluene. $^a$ Experimental absorption maxima from Ref. \cite{gho15}}
\end{table}

\section{Conclusions}
\label{conclusions}
The S$_1 \leftarrow$ S$_0$ absorption energies of three different forms of the gas phase oxyluciferin have been explored by means of high-level methods such as the VMC for the ground state geometry and the evGW/BS formalism for the excitation energies. 
Unlike findings shown for other biological chromophores, like the protonated Schiff base of the retinal and the peridinin carotenoid \cite{Coccia:2012ex,Coccia14}, the accuracy in the determination of the ground-state geometry of keto-1 and enol-1 forms affects the value of the vertical excitation S$_1 \leftarrow$ S$_0$ energy only slightly.  In the case of OxyLuc forms, the optimized ground state structure plays a minor role in the determination of the optical properties, with only a moderate blue shift in the vertical absorption when the difference between single and double bonds increases, as for CASSCF(18,15) structures.
The value of 2.32 eV obtained by applying the BS equation on the VMC structure of the keto-1 form shows an excellent agreement with the experimental 2.26(8) eV finding.

As shown in previous works, VMC is able to include a balanced description of the electronic correlation. Assuming the VMC as a reference structure, we have observed that the much less computationally demanding CAM-B3LYP  or M062X DFT structural relaxation provides accurate results for complex, non-symmetric and conjugated molecular systems as the anionic forms of OxyLuc. 

Regardless of the level of theory employed for the ground-state geometry optimization, the keto-1/enol-1 shift computed at BS level has the same sign of the very accurate MS-CASPT2 results while TDDFT calculations using semi-local functionals and even hybrids as B3LYP fail to predict the correct sign of the shift. However, when passing from LDA to CAM-B3LYP wave functions in evGW/BS calculations a shift of 0.1 eV is observed for the enol-1 form. This indicates that particular care needs to be paid in the choice of the initial guess when determining the keto-1/enol-1 shift. 

In the last years the BS approach has been shown to provide accurate excitation energies on well-known sets of  small molecules and organic dyes. The present work contributes, together with previous ones by us and other authors~\cite{ma2009modeling, yin2014charge, faber2013many}, in showing that it is a promising tool to compute excited state energies in complex chromophores of biological interest.

 Finally, studying the dianonic enol-2 form we have shown that particular attention needs to be paid to its unbound electronic character, and that meaningful results are obtained only when considering the effect of stabilizing solvents.
 
Furthermore, vibronic terms could have a not negligible weight in the applied theoretical model and a deeper and more quantitative insight on these effects on the vertical absorption of the OxyLuc forms requires further investigation. 
\begin{acknowledgement}
EC thanks the Labex MiChem part of French state funds managed  by  the  ANR  within  the  Investissements  d'Avenir  programme (contract grant number: ANR-11-IDEX-0004-02), the computational centre of the Laboratoire de Chimie Th\'eorique and the University of L'Aquila (SIR fellowship) for partial financial support. 
DV acknowledges partial support from EU Centre  of Excellence MaX-MAterials at the eXascale (H2020 grant no. 676598). Computational resources were provided by the Caliban computer Centre of the University of L'Aquila, the Hydra cluster of the computational centre of the University of Modena and Reggio Emilia, and by the PRACE research infrastructure (project BIOCHROMO). Authors thank S. Caprasecca for the careful reading of the manuscript.
This work has been supported by the European Research Council Project MultiscaleChemBio
(n. 240624) within the VII Framework Program of the European Union.
\end{acknowledgement}

 \begin{suppinfo}
Table S1 contains the optimized angles and dihedrals (in degrees) of the DFT, VMC and CASSCF(18,15) structures of  keto-1, enol-1 and  enol-2 forms. 
TDDFT values on various structures of keto-1 and enol-1 are reported in Tables S2 and S3. Convergence analysis of the absorption energy with respect to the size of the basis set is shown in Table S4 (TDDFT), S5 (BS) and S6 (CC2). The evGW/BS excitations for keto-1 and enol-1 starting from an LDA ground state are reported in Table S7. HOMO, LUMO energies and gaps  of gas-phase keto-1, enol-1 and enol-2, using various functionals and basis sets, are given in Table S8. In Table S9  the same quantities for enol-2 calculated using plane-wave basis sets are given. In Tables S10, S11 and S12 HOMO, LUMO energies and gaps in water and toluene  are collected for enol-2, keto-1 and enol-1, respectively.  
 \end{suppinfo}

\bibliography{bib}
\end{document}